\documentclass{aa}  

\usepackage{graphicx}
\usepackage{txfonts}
\usepackage[]{hyperref} 
\usepackage{booktabs} 
\usepackage{array} 
\usepackage{multirow} 
\usepackage{paralist} 
\usepackage{verbatim} 
\usepackage{ulem}
\usepackage{xspace}
\usepackage{siunitx}
\usepackage{amssymb}
\usepackage{capt-of}
\usepackage{natbib}
\bibpunct{(}{)}{;}{a}{}{,} 

\graphicspath{{./img/}}

\makeatletter
\renewcommand*\aa@pageof{, page \thepage{} of \pageref*{LastPage}}
\makeatother

\addto\extrasenglish{%
}

\newcommand\freefootnote[1]{%
  \let\thefootnote\relax%
  \footnotetext{#1}%
  \let\thefootnote\svthefootnote%
}

\begin{document}

   \title{Upper limits on atmospheric abundances of KELT-11b and WASP-69b from a retrieval approach}

   \subtitle{}

   \author{
            F.~Lesjak \inst{1,2}\thanks{\email{flesjak@aip.de}}\and 
            L.~Nortmann \inst{2} \and
            D.~Cont \inst{3,4}  \and
            P.\,J.~Amado \inst{5}  \and
            M. Azzaro \inst{6}\and
            J.\,A.~Caballero \inst{7}  \and
            S. Czesla \inst{8} \and
            A.~Hatzes \inst{8}  \and
            Th.~Henning \inst{9}  \and
            M. López-Puertas \inst{5} \and
            K. Molaverdikhani \inst{3,4} \and
            D. Montes \inst{10}\and
            J. Orell-Miquel \inst{11} \and
            E.~Pall\'e \inst{12,13}  \and
            A.~Peláez-Torres \inst{5} \and
            A.~Quirrenbach \inst{14}  \and
            A.~Reiners \inst{2}  \and
            I.~Ribas \inst{15,16}  \and
            A. S\'anchez-L\'opez \inst{5} \and
            A.~Schweitzer \inst{17}  \and
            F. Yan \inst{18}
          }

   \institute{
Leibniz Institute for Astrophysics Potsdam (AIP), An der Sternwarte 16, 14482 Potsdam, Germany \and
            Institut f\"ur Astrophysik und Geophysik, Georg-August-Universit\"at, Friedrich-Hund-Platz 1, 37077 G\"ottingen, Germany \and
            Universitäts-Sternwarte, Fakultät für Physik, Ludwig-Maximilians-Universität München, Scheinerstr. 1, 81679 München, Germany \and
            Exzellenzcluster Origins, Boltzmannstraße 2, 85748 Garching, Germany \and
            Instituto de Astrofísica de Andalucía (IAA-CSIC), Glorieta de la
            Astronomía s/n, 18008 Granada, Spain \and
            {Centro Astron\'omico Hispano en Andaluc\'ia, 
        Observatorio Astron\'omico de Calar Alto, 
        Sierra de los Filabres, 04550 G\'ergal, Almer\'ia, Spain} \and
            {Centro de Astrobiolog\'ia (CSIC-INTA), 
        Camino Bajo del Castillo s/n, Campus ESAC, 
        28692 Villanueva de la Ca\~nada, Madrid, Spain} \and
            Thüringer Landessternwarte Tautenburg, Sternwarte 5, 07778
            Tautenburg, Germany \and
            Max-Planck-Institut für Astronomie, Königstuhl 17, 69117 Heidelberg, Germany \and
            Departamento de Física de la Tierra y Astrofísica and IPARCOSUCM (Instituto de Física de Partículas y del Cosmos de la UCM), Facultad de Ciencias Físicas, Universidad Complutense de Madrid, 28040, Madrid, Spain \and
            Department of Astronomy, University of Texas at Austin, 2515 Speedway, Austin, TX 78712, USA \and
            Instituto de Astrofísica de Canarias (IAC), Calle Vía Lactea s/n, 38200 La Laguna, Tenerife, Spain \and
            Departamento de Astrofísica, Universidad de La Laguna, 38026 La Laguna, Tenerife, Spain \and
            Landessternwarte, Zentrum für Astronomie der Universität Heidelberg, Königstuhl 12, 69117 Heidelberg, Germany \and
            Institut de Ciències de l’Espai (CSIC-IEEC), Campus UAB, c/ de Can Magrans s/n, 08193 Bellaterra, Barcelona, Spain \and
            Institut d’Estudis Espacials de Catalunya (IEEC), 08034 Barcelona, Spain \and
            Hamburger Sternwarte, Gojenbergsweg 112, 21029 Hamburg, Germany \and
            Department of Astronomy, University of Science and Technology of China, Hefei 230026, People’s Republic of China
             }

   \date{Received 03 June 2025 / Accepted 22 October 2025}

  \abstract
   {WASP-69\,b and KELT-11\,b are two low-density hot Jupiters, which are expected to show strong atmospheric features in their transmission spectra. Such features offer valuable insights into the chemical composition, thermal structure, and cloud properties of exoplanet atmospheres. High-resolution spectroscopic observations can be used to study the line-forming regions in exoplanet atmospheres and potentially detect signals despite the presence of clouds.}
   {We aimed to detect various molecular species and constrain the chemical abundances and cloud deck pressures using high-resolution spectroscopy.}
   {We observed multiple transits of these planets with CARMENES and applied the cross-correlation method to detect atmospheric signatures. Further, we used an injection-recovery approach and retrievals to place constraints on the atmospheric properties.}
   {We detected a tentative H$_2$O signal for KELT-11\,b but not for WASP-69\,b, and searches for other molecules such as H$_2$S and CH$_4$ resulted in non-detections for both planets. By investigating the signal strength of injected synthetic models, we constrained which atmospheric abundances and cloud deck pressures are consistent with our cross-correlation results. In addition, we show that a retrieval-based approach leads to similar constraints of these parameters.}
   {}


   \maketitle

\section{Introduction}
Hot, low-density gas giants offer unique opportunities to probe atmospheric composition and structure using transmission spectroscopy techniques. Their inflated atmospheres result in strong transmission signals, which can be studied using high-resolution Doppler spectroscopy. Two such planets are WASP-69\,b and KELT-11\,b, which have very similar sizes but different equilibrium temperatures.

KELT-11\,b is a hot sub-Saturn with an equilibrium temperature of about 1700\,K \citep{pepper_kelt-11b_2017} on a 4.7\,d orbit around its G-type host star. The atmosphere of this highly inflated planet was first analysed by \citet{zak_high-resolution_2019}, who found no evidence of sodium absorption in the transmission spectrum observed with the High Accuracy Radial Velocity Planet Searcher (HARPS). A follow-up study by \citet{mounzer_hot_2022} using an updated transit ephemeris did detect a sodium feature, but the signal was found to be weaker than anticipated based on comparisons with other sub-Saturn planets. The low signal strength was interpreted as potential evidence of a significant cloud coverage muting the spectral features of KELT-11\,b's atmosphere. In contrast, \citet{colon_unusual_2020} and \mbox{\citet{changeat_kelt-11_2020}} detected the presence of water using a combination of \textit{Hubble} Space Telescope (HST) and Transiting Exoplanet Survey Satellite (TESS) data, although they report an unusually shaped absorption feature and a low H$_2$O abundance ($\log_{10}(\mathrm{H}_2\mathrm{O}) = -5.9 ^{+0.2}_{-0.4}$ for their base retrieval). An additional absorption feature could be attributed to a mix of carbon-bearing species, and their retrievals are unable to place tight constraints on the presence of clouds. These unusual results make KELT-11\,b an interesting target for analysing the atmospheric content of H$_2$O and carbon-bearing molecules, which until now had not been conducted with high-resolution spectroscopy.

WASP-69\,b, an inflated Saturn-mass planet with an equilibrium temperature of 963\,K orbiting a K-type star \citep{anderson_three_2014}, has been the subject of extensive atmospheric investigations, which have revealed a complex and dynamic exoplanetary environment: High-resolution spectroscopic studies have consistently detected sodium in its atmosphere, and a disparity between the amplitudes of the D1 and D2 lines is potentially caused by the presence of atmospheric aerosols \citep{casasayas-barris_detection_2017, khalafinejad_probing_2021, langeveld_survey_2022, sicilia_gaps_2025}.
An outflow of an extended helium tail from WASP-69\,b's atmosphere was discovered by \citet{nortmann_ground-based_2018}. \citet{tyler_wasp-69bs_2024} further characterised this outflow, revealing that the escaping envelope extends up to 7.5 planetary radii behind the planet. The strong helium feature was additionally confirmed in studies by \citet{vissapragada_upper_2022}, \citet{allart_homogeneous_2023}, \citet{levine_exoplanet_2024}, \citet{guilluy_gaps_2024}, and \citet{masson_probing_2024}. The first molecular feature detected was H$_2$O absorption, observed with the \textit{Hubble} Wide Field Camera 3 (WFC3) by \citet{tsiaras_population_2018}. \citet{guilluy_gaps_2022} employed a high-resolution cross-correlation method and claimed the detection of CO, CH$_4$, NH$_3$, H$_2$O, and C$_2$H$_2$ in WASP-69\,b's atmosphere. The presence of these species, coupled with the notable absence of CO$_2$, suggests a high C/O ratio and disfavours models with a metallicity greater than 10 times the solar value. Absorption features of H$_2$O and NH$_3$ and evidence of a high-altitude cloud deck were also detected by \citet{petit_dit_de_la_roche_detection_2024} using the low-resolution spectrographic mode of the FOcal Reducer/low dispersion Spectrograph 2 (FORS2) at the Very Large Telescope.
Complementing these molecular detections, optical and near-infrared transmission spectroscopy has unveiled a strong Rayleigh scattering slope, indicative of high-altitude aerosols \citep{murgas_gtc_2020, estrela_detection_2021, ouyang_tentative_2023}. Secondary eclipse measurements from \textit{Spitzer} suggest the absence of a temperature inversion in WASP-69\,b's atmosphere \citep{baxter_transition_2020} and hint at a potentially high atmospheric metallicity exceeding 30 times the solar value \citep{wallack_investigating_2019}. \citet{schlawin_multiple_2024} observed two secondary eclipses with the \textit{James Webb} Space Telescope (JWST) to measure the emission spectrum from 2-12\,$\mu$m, which shows features of H$_2$O, CO, and CO$_2$. In addition, strong scattering or significant cloud coverage is necessary to achieve a good fit to the data.

In this work we analysed ground-based transmission spectra of KELT-11\,b and WASP-69\,b and find no conclusive evidence of atmospheric absorption in either case. Nevertheless, we place tight constraints on the molecular abundances and cloud deck heights.

\section{Observations and data reduction} \label{Sect_Observations}

\begin{table*}[ht]\renewcommand{\arraystretch}{1.5}
\centering 
 \caption[]{Stellar and planetary parameters of the WASP-69 and KELT-11 systems.}\label{Table_Parameters}
\begin{tabular}{llclcl}
 \hline \hline
  Parameter &
  Symbol &
  WASP-69\,b &
  Reference &
  KELT-11\,b &
  Reference
 \\ \hline
\textit{Planet}   &   & & & &\\
Radius   & $R_\mathrm{p}$ ($R_\mathrm{Jup}$) &  $1.11 \pm 0.04$ & (1) & $1.35 \pm 0.10$ & (7)\\
Mass   & $M_\mathrm{p}$ ($M_\mathrm{Jup}$) &  $0.29 \pm 0.03$ & (1) & $0.171 \pm 0.015$ & (7)\\
Orbital period   & $P_\mathrm{orb}$ (days) & $3.868140 \pm 0.000002$  & (1) & $4.7362006 \pm 0.0000034$ & (8)\\
Orbital inclination    & $i$ ($^\circ$) &  $86.71 \pm 0.20$ & (1) & $85.3 \pm 0.2$ & (7)\\
Orbital eccentricity   & $e$ & 0.0 & (1) & $0.0007^{+0.00200}_{-0.00050}$ & (7)\\
Argument of periastron & $\omega$ ($^\circ$) & 90 & (2) & $-1^{+147.4}_{-74.6}$ & (7) \\
Semi-major axis   & $a$ (au)& $0.04525 \pm 0.00075$  & (2)& $0.0623 \pm 0.0006$ & (7)\\
Time of mid-transit   & $T_0$ (BJD$_\mathrm{TDB}$) & 2459798.7777552 & (3) & 2458255.43247 & (8)\\
RV semi-amplitude   & $K_p$ (km\,s$^{-1}$) & 127.1 $\pm$ 2.1 & (4)& 142.6 $\pm$ 1.4 & (4)\\
Surface gravity   & $\log{g}$ (cgs)& 2.79 $\pm$ 0.04 & (5) & 2.39 $\pm$ 0.04 & (5)\\
Equil. temperature   & $T_\mathrm{eq}$ (K)& $963 \pm 18$ & (2) & $1712^{+51}_{-46}$ & (9)\\
\noalign{\smallskip}
\hline
\textit{Star}   &   &  & & &\\
Radius   & $R_\star$ ($R_\odot$) & 0.86 $\pm$ 0.03 & (1) & $2.69 \pm 0.04$ & (7)\\
Effective temperature   & $T_\mathrm{eff}$ (K)& 4700 $\pm$ 500 & (1) & $5375 \pm 25$ & (7)\\
Systemic velocity   & $v_\mathrm{sys}$ (km\,s$^{-1}$) & $-9.63$ & (6) & $35.0 \pm 0.1$& (9)\\
$J$-band magnitude   & $J_s$ (mag) & 8.03 & (10) & 6.62 & (10)
 \\ \hline
\end{tabular}
\tablebib{
(1) \citet{stassun_accurate_2017}, (2) \citet{casasayas-barris_detection_2017}, (3) \citet{saha_precise_2023}, (4) Calculated from orbital parameters: $K_\mathrm{p} = 2\,\pi\,a\,\sin(i)\,P^{-1}\,(1-e^2)$, (5) Calculated from planetary parameters: $\log{g} = \log({GM_\mathrm{p}/R_\mathrm{p}^2})$, (6) \citet{anderson_three_2014}, (7) \citet{beatty_determining_2017}, (8) \citet{kokori_exoclock_2023}, (9) \citet{pepper_kelt-11b_2017}, (10) \citet{skrutskie_two_2006}
}
\end{table*}

\begin{table*}\renewcommand{\arraystretch}{1.5}
\caption{Observational conditions for the six nights of WASP-69\,b and two nights of KELT-11\,b observations.}  
\label{table:Night Properties}    
\centering                        
\begin{tabular}{c c c c c c c c}  
\hline\hline  
Night & Date & Exposure time ($s$) & \textit{N}$_\mathrm{exp}$ &Mean S/N & Airmass (start - mid - end) & Rel. humidity (\%) \\ 
\hline  
\textit{WASP-69\,b}   & &  & & & &\\
1 & 2017-08-22 & 395 & 35 & 59 & 1.46 - 1.35 - 1.97 & 51 \\  
2 & 2017-09-22 & 398 & 31 & 54 & 1.39 - 1.35 - 1.97 & 61 \\  
3 & 2020-08-13 & 398 & 34 & 59 & 1.94 - 1.35 - 1.52 & 48 \\  
4 & 2021-08-27 & 396 & 25 & 47 & 1.36 - 1.35 - 2.02 & 46 \\  
5 & 2022-08-14 & 392 & 20 & 51 & 1.51 - 1.35 - 1.38 & 45 \\  
6 & 2023-07-01 & 396 & 28 & 63 & 1.97 - 1.35 - 1.36 & 79 \\  
\noalign{\smallskip}
\hline
\textit{KELT-11\,b}   &  & & & & &\\
1 & 2023-01-10 & 306 & 51 & 95 & 1.97 - 1.46 - 1.92 & 44 \\  
2 & 2023-03-27 & 306 & 49 & 71 & 1.84 - 1.46 - 2.02 & 39 \\  
\hline  

\end{tabular}
\end{table*}

Table \ref{Table_Parameters} presents the stellar and planetary parameters adopted in this work.
We observed the transmission spectra of WASP-69\,b and \mbox{KELT-11\,b} with the CARMENES spectrograph at the Calar Alto Observatory \citep{quirrenbach_carmenes_2014, quirrenbach_carmenes_2020}. KELT-11\,b was observed during two nights in January and March of 2023, and WASP-69\,b during six nights between August 2017 and July 2023. The dates and observational conditions for all nights are summarised in Table \ref{table:Night Properties} and shown in Fig. \ref{Fig_ObservationalConditions}.
The observations of WASP-69\,b include full transits from ingress to egress and additional out-of-transit exposures at the beginning and end.
KELT-11\,b has a long transit duration of 7.2\,h, preventing the observation of a full transit with an out-of-transit baseline from the Calar Alto Observatory. Consequently, neither of the two nights cover the entire transit. In this work, we analysed the data from the near-infrared channel with a wavelength coverage of $9\,600-17\,100\,\AA$ and a resolution of $R\approx80\,400$. The molecular species with the highest expected abundances (H$_2$O, H$_2$S, and CH$_4$) have strong spectral lines in this wavelength region.

CARMENES has two input fibres, and we observed the targets with fibre A while fibre B was used to measure the sky background. The reduction pipeline {\tt caracal} v2.00 \mbox{\citep{zechmeister_flat-relative_2014, caballero_carmenes_2016}} was used to reduce the raw frames for the extraction of the 1D spectra of each order with corresponding uncertainties and flux signal-to-noise ratio (S/N). Due to strong telluric water absorption features, the spectral regions at 1.10--1.18\,$\mu$m (echelle orders 55--52) and 1.34--1.50\,$\mu$m (orders 45--41) have low flux levels and were excluded from our analysis.

\subsection{Normalisation} \label{Section_Normalisation}

Our spectral normalisation procedure involved three main steps: removing outliers, adjusting for continuum variations, and masking telluric and sky emission lines.
To remove outliers, we analysed the time series of each pixel. Any data points that deviated by more than 5$\sigma$ were removed and replaced using linear interpolation. In cases where more than three values in a pixel's time series required replacement, we opted to mask that pixel entirely across all spectra in the time series.

To correct for continuum flux variations, we used the method described in \citet{lesjak_retrieval_2023} to identify points that follow the continuum's shape, which can then be used for polynomial fitting. We began by creating a master spectrum for each order by averaging all exposures over time. This master spectrum was divided into 100 equal-sized wavelength bins. For each bin, we selected the 90\% percentile value of all flux values within it. These 100 points were further consolidated into ten bins, from which we chose the second largest value in each bin. The resulting ten points effectively captured the overall shape of the blaze function without being influenced by individual telluric emission or absorption lines. We then fitted a third-degree polynomial to these ten data points and divided all spectra and errors by this fit.

Following this, we further divided each individual spectrum by a linear fit to ensure a consistent continuum level across all spectra. We completely masked deep telluric lines that fell below 40\% of the continuum flux at any time of the observation. In addition, one pixel adjacent to the left and right of these masked regions was also excluded.
As a final step, sky emission lines with a flux exceeding 107\% of the continuum were masked.

\subsection{Removal of stellar and telluric lines} \label{Sect_RemovingTellurics}

\begin{figure}
\centering
\includegraphics[width=\hsize]{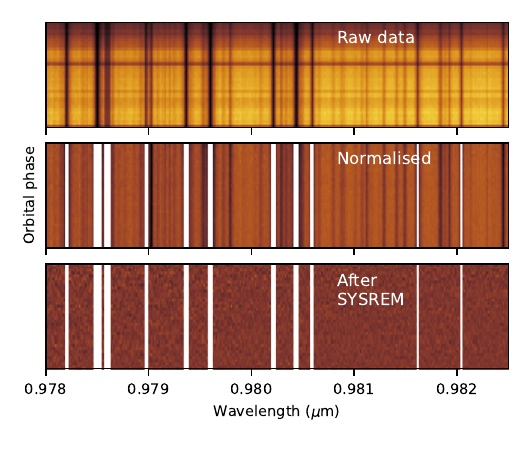}
  \caption{Data reduction steps of a representative wavelength range for night 1 of WASP-69\,b. Top panel: Unprocessed spectra as produced by the CARACAL pipeline. Middle panel: After normalisation to bring all spectra onto a common continuum level, and masking of deep telluric lines. Bottom panel: After the removal of stellar and telluric lines with \texttt{SYSREM}.
          }
     \label{Fig_DataReduction}
\end{figure}

For our data processing, we used \texttt{SYSREM} \citep{tamuz_correcting_2005, birkby_detection_2013} to remove telluric and stellar lines from the spectra. \texttt{SYSREM} employs a principal component analysis to model the linear components across wavelength and time, while accounting for individual data point uncertainties. Stellar, telluric, and planetary components exhibit different Doppler shifts in the observed spectral time series. Stellar and telluric lines remain relatively static in velocity space throughout the observation, while the planetary lines shift significantly due to the change in radial velocity during the transit.
Following the approach of \citet{gibson_relative_2022}, we first divided each order by its time-averaged spectrum. Subsequently, we applied \texttt{SYSREM} to iteratively remove linear trends in the time and spectral domain, effectively eliminating the quasi-static stellar and telluric components. We refined the linear model in each iteration until the average relative change dropped below 0.01. This threshold ensures a sufficient convergence while maintaining a reasonable runtime. The model was then subtracted from the data before proceeding to the next iteration. This process resulted in residuals primarily composed of noise and the potential planetary signal, which experiences wavelength shifts over time and is thus not as readily modelled by \texttt{SYSREM} as the quasi-static telluric and stellar components. Figure \ref{Fig_DataReduction} illustrates an example of the resulting residual spectra.
It is worth noting that the number of \texttt{SYSREM} iterations can influence the strength of the recovered signal \citep[e.g.][]{alonso-floriano_multiple_2019,cabot_robustness_2019,boldt-christmas_optimising_2024}. To determine the optimal number of iterations, we initially followed the injection-based approach of \citet{cheverall_robustness_2023} as summarised in the following: A planetary model spectrum was first Doppler-shifted using the expected $K_\mathrm{p}$ and $v_\mathrm{sys}$ values and injected into the raw data. The data were pre-processed and cross-correlated to yield the injected cross-correlation function (CCF$_\mathrm{inj}$). The differential cross-correlation between the data with and without an injected model, $\Delta$CCF = CCF$_\mathrm{inj}$ $-$ CCF, was then computed and converted into a S/N map for each iteration, combining the information of all spectral orders. We selected the iteration with the highest S/N significance as the optimal number of \texttt{SYSREM} iterations.

This procedure was carried out for each combination of night and species separately, and in the following we used the iterations summarised in Table \ref{table:Sysrem iterations} for the analysis of KELT-11\,b. For WASP-69\,b, however, strong telluric residuals remained in the cross-correlation function (CCF) after applying the number of iterations as optimised with this method (Fig. \ref{Fig_CCF_W69_optimised}). Because the planetary rest frame overlaps with the telluric rest frame for significant parts of the transit during most nights (see Fig. \ref{Fig_ObservationalConditions}), a masking of the affected regions would have resulted in a significant loss of in-transit information. Instead, we opted for increasing the number of \texttt{SYSREM} iterations to mitigate the telluric contamination. For each night, we looked at the relative change of the standard deviation of the \texttt{SYSREM}-subtracted residual data between subsequent iterations. This metric reaches a plateau once the most significant systematics are removed (Spring \& Birkby, in prep.). Similar to the previous method, this approach also results in a relatively low number of iterations (between three and five iterations for the different nights). However, visual inspection still showed significant telluric contamination in the CCF plots. Therefore, we opted to use a more conservative approach by applying seven iterations across all nights and species, which led to a significant reduction of the disruptive residuals on most nights (Fig. \ref{Fig_CCF_W69}). Varying the number of iterations between five and eight did not lead to any major differences in the change of the standard deviation. This approach aimed to mitigate biases from inadequately removed residuals and to avoid misinterpreting spurious features as tentative detections. However, applying more \texttt{SYSREM} iterations also removes a greater portion of the planetary signal, potentially leading to more conservative upper limits in the subsequent analysis.
Night 3 showed strong telluric residuals that could not be alleviated even with these additional passes of \texttt{SYSREM}. Because these contaminations affect almost all of the in-transit spectra, we excluded this night from the following analysis.

\section{Analysis and results} \label{Sect_Analysis}
\subsection{Cross-correlation}

\subsubsection{Model spectra}
We used \texttt{petitRADTRANS} \citep{molliere_petitradtrans_2019} to generate synthetic model spectra for the cross-correlation analysis. We chose isothermal temperature-pressure profiles fixed to the equilibrium temperature of the two planets (963\,K for WASP-69\,b and 1712\,K for KELT-11\,b). We assumed isobaric abundances determined by one abundance value for each species, and the opacities were included using the line lists of \citet{polyansky_exomol_2018} for H$_2$O, \citet{rothman_hitran2012_2013} for H$_2$S, and \citet{hargreaves_accurate_2020} for CH$_4$. The effect of clouds was modelled using the grey-cloud model from \mbox{\texttt{petitRADTRANS}}, which assumes a completely opaque cloud layer blocking any contributions from layers below a certain pressure level $P_\mathrm{cloud}$. We further included continuum absorption of H$^-$ and collision-induced absorption of H$_2$-H$_2$ and \mbox{H$_2$-He} \mbox{\citep[and references therein]{borysow_collision-induced_2002, richard_new_2012}}. \texttt{petitRADTRANS} calculates the wavelength-dependent planetary radius $R_p$, which we converted into the transmission spectrum $T(\lambda) = 1 - ({R_{p}(\lambda)}/{R_{*}})^2$. Subsequently, the spectra were convolved to the instrumental resolution of the CARMENES near-infrared spectrograph ($R \approx 80\,400$) using a Gaussian instrumental profile. We approximated the pseudo-continuum of the model by applying a 99th-percentile filter with a width of 5000 spectral bins, followed by a Gaussian filter with a width of 500 bins. Finally, the model was divided by this pseudo-continuum approximation.

\subsubsection{Cross-correlation method}
In our analysis, we used a weighted CCF to extract the planetary signal from the residual spectra, following the methodology from \citet{gibson_detection_2020} and \mbox{\citet{cont_atmospheric_2022}}:
\begin{align}
    \mathrm{CCF}(v, t) = \sum _{i = 0}^N \frac{R_i(t) \cdot M_i(v)}{\sigma_{i}(t)^2}\,,
\end{align}
where $t$ and $i$ are the time and pixel indices, respectively, $R$ represents the residual spectra with associated uncertainties $\sigma$, and $M$ is the model spectrum. For the cross-correlation analysis, we used individual models for each species and fixed the volume mixing ratio (VMR) to $-3$. of We applied Doppler shifts to the model spectrum over a velocity range of $-1000$\,km\,s$^{-1}$ to $+1000$\,km\,s$^{-1}$ in steps of $1.3$\,km\,s$^{-1}$ and calculated the cross-correlation value for each shifted model. This process generated a 2D CCF map for each spectral order, which were then combined by calculating the mean CCF (Figs. \ref{Fig_CCF_K11} and \ref{Fig_CCF_W69}).

To combine the information of multiple nights, we merged the individual CCFs into a single array, from which we subsequently constructed a $K_\mathrm{p}$-$v_\mathrm{offset}$ map. This involved exploring various orbital semi-amplitude ($K_p$) values and shifting the CCF into the corresponding planetary reference frames according to the Doppler velocity:
\begin{align}
v_p = K_p \cdot \left[ \cos(\nu (\phi) + \omega) + e \cdot \cos(\omega) \right]  + v_\mathrm{sys} - v_\mathrm{bary} + v_\mathrm{offset}\,,
\label{Eqvelocity}
\end{align}
where $\nu (\phi)$ represents the true anomaly at an orbital phase $\phi$, $e$ is the orbital eccentricity, $\omega$ the argument of periastron, $v_\mathrm{sys}$ and $v_\mathrm{bary}$ are the systemic and barycentric velocities, respectively, and $v_\mathrm{offset}$ accounts for potential deviations from the planetary rest frame. In practice, we calculated $K_\mathrm{p} = 2\,\pi\,a\,\sin(i)\,P^{-1}\,(1-e^2)$ and used the \texttt{kepler.rv\_drive} method from the python package \texttt{radvel}\footnote{\url{https://radvel.readthedocs.io/en/latest/}} \citep{fulton_radvel_2018} to calculate the true anomaly and determine the radial velocity. We averaged the shifted CCFs along the time axis, resulting in 1D vectors, and stacked these vectors for each trial $K_p$ value (in the range from 0\,km\,s$^{-1}$ to 400\,km\,s$^{-1}$ in steps of 1\,km\,s$^{-1}$) to create the 2D $K_\mathrm{p}$-$v_\mathrm{offset}$ map. This map was converted into S/N units by dividing it by the standard deviation calculated from regions distant from the expected signal location ($|v| > 50$\,km\,s$^{-1}$).

\subsubsection{Signals}
\begin{figure*}
\centering
\includegraphics[width=\hsize]{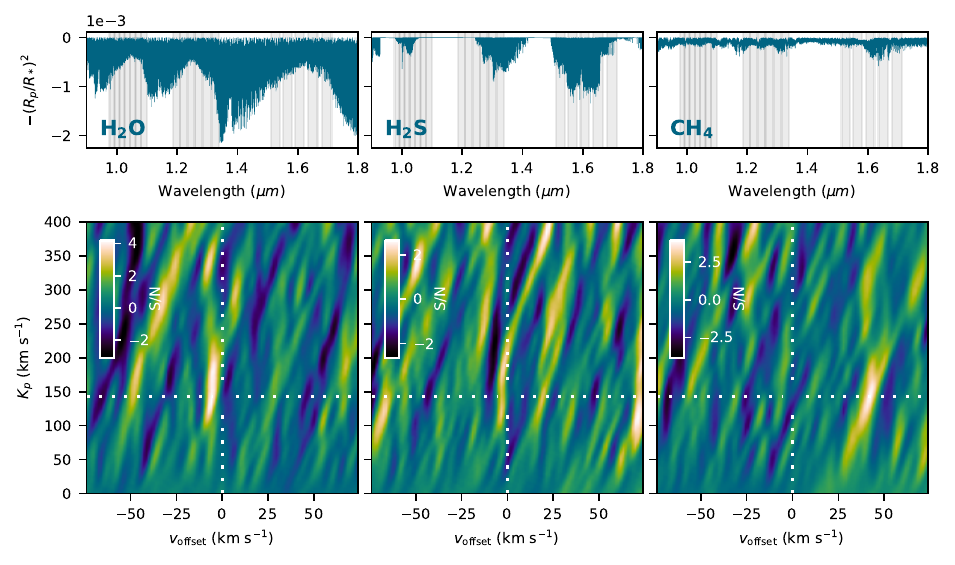}
  \caption{Model spectra (top) and $K_\mathrm{p}$-$v_\mathrm{offset}$ maps for the combination of all nights (bottom) of H$_2$O, H$_2$S, and CH$_4$ for KELT-11\,b. The grey shaded areas show the wavelength regions of the spectral orders used for the analysis. The dotted lines indicate the expected location of the planetary signal.
          }
     \label{Fig_Models+Detmaps_K11}
\end{figure*}

For both planets, we searched for atmospheric absorption signals of H$_2$O, H$_2$S, and CH$_4$. For KELT-11\,b, we determined the optimal number of \texttt{SYSREM} iterations as explained in Sect. \ref{Sect_RemovingTellurics} and detailed in Table \ref{table:Sysrem iterations}.
We found a tentative signal of H$_2$O with a S/N of 4.1 close to the literature $K_\mathrm{p}$ value of 143\,km\,s$^{-1}$ and blue-shifted from the expected planetary rest frame by 6\,km\,s$^{-1}$ (Fig. \ref{Fig_Models+Detmaps_K11}). However, the signal strength of H$_2$O was not sufficient to firmly confirm its planetary origin, and there was no clear trail in the CCF map (Fig. \ref{Fig_CCF_K11}). The $K_\mathrm{p}$-$v_\mathrm{offset}$ maps of the individual nights show that this tentative signal mostly originates from night 1 (Fig. \ref{Fig_Detmap_K11_IndividualNights}). On this night, the radial velocity separation between the planetary and telluric reference frames was small during the transit (Fig. \ref{Fig_ObservationalConditions}), which exacerbated the telluric contamination. None of the other species were detectable, and we note that all spurious peaks in the $K_\mathrm{p}$-$v_\mathrm{offset}$ maps were weaker than the tentative H$_2$O signal.

In the case of WASP-69\,b, we initially also determined the optimal number of iterations following the injection-based method in Sect. \ref{Sect_RemovingTellurics}, resulting in 3, 2, 6, 3, 2, and 2 iterations for H$_2$O in nights 1 through 6, respectively. The corresponding CCFs and the $K_\mathrm{p}$-$v_\mathrm{offset}$ map for H$_2$O are shown in Figs. \ref{Fig_CCF_W69_optimised} and \ref{Fig_Detmap_W69_H2O_OptimisedSYSREM}, respectively. While this resulted in a tentative signal with a S/N of 4.1 close to the expected position in velocity space, the CCFs show significant telluric contamination on most of the nights. The relatively small number of \texttt{SYSREM} iterations on most nights led to the strongest recovery of an injected signal, but did not sufficiently remove the telluric lines. On most of the observed nights, the planetary trail crosses the location of the telluric contamination (at $v = 0$\,km\,s$^{-1}$), so that a simple masking of the affected region is not possible. As discussed in Sect. \ref{Sect_RemovingTellurics}, we therefore decided to increase the number of \texttt{SYSREM} iterations and applied seven iterations to all of the nights. This led to a significant reduction of the contamination in the CCFs (Fig. \ref{Fig_CCF_W69}).

\begin{figure*}
\centering
\includegraphics[width=\hsize]{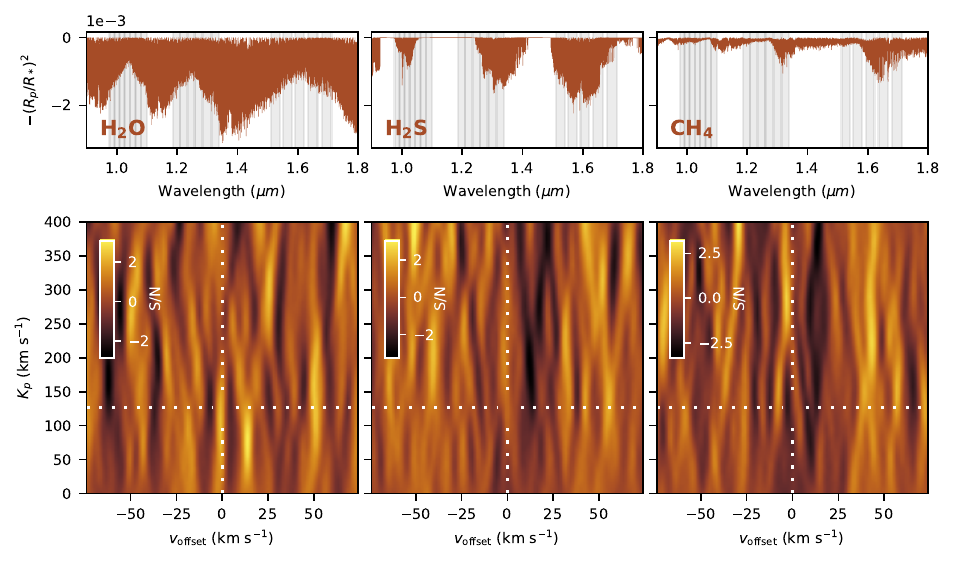}
  \caption{Same as Fig. \ref{Fig_Models+Detmaps_K11} but for WASP-69\,b.
          }
     \label{Fig_Models+Detmaps_W69}
\end{figure*}

Figure \ref{Fig_Models+Detmaps_W69} shows the $K_\mathrm{p}$-$v_\mathrm{offset}$ maps of H$_2$O, H$_2$S, and CH$_4$ for WASP-69\,b after seven \texttt{SYSREM} iterations and the exclusion of night 3.
The tentative H$_2$O signal was now reduced in signal strength and was of a similar amplitude as the surrounding noise pattern. This prevented us from establishing a firm detection of H$_2$O in this planet.
In addition, we report non-detections for H$_2$S and CH$_4$.

The non-detections in both planets indicate that the strengths of the absorption lines are not sufficient to be confidently detectable with our observations. A weak signal can be caused either by low abundances or by an extensive cloud deck blocking the contribution of deeper atmospheric layers. In the following, we place constraints on the abundances and the cloud deck height based on these non-detections, and investigate which H$_2$O abundances would be consistent with the observed tentative signal if it was of planetary origin.

\begin{table}\renewcommand{\arraystretch}{1.5}
\caption{Number of \texttt{SYSREM} iterations used in the analysis.}             
\label{table:Sysrem iterations}      
\centering                          
\begin{tabular}{c c c c c}        
\hline\hline                 
 & Night & \multicolumn{3}{c}{$N$} \\
 & & H$_2$O & H$_2$S & CH$_4$  \\    
\hline                        
    KELT-11\,b & 1 & 4 & 3 & 3 \\ 
    & 2 & 3 & 4 & 4 \\ 
    WASP-69\,b & all & 7 & 7 & 7\\
\hline 
\end{tabular}
\tablefoot{See Sect. \ref{Sect_RemovingTellurics} for details on the selection process.} 
\end{table}

\subsection{Constraints on abundances and cloud height}
To determine upper limits on the abundances and the cloud deck pressure of the planets' atmospheres, we employed an injection-recovery test based on the S/N in the $K_\mathrm{p}$-$v_\mathrm{offset}$ map \citep[see e.g.][]{grasser_peering_2024, parker_limits_2025}.

\subsubsection{S/N grid}
To determine whether a species in an atmosphere with given chemical abundances and cloud deck height should have been detectable in the cross-correlation analysis, we first injected a synthetic model spectrum with these parameters into the observed data. For each exposure, the model was shifted to the rest frame of the tentative H$_2$O detection (using Eq. \ref{Eqvelocity} with $K_\mathrm{p} = 89$\,km\,s$^{-1}$ and $v_\mathrm{offset}=-0.8$\,km\,s$^{-1}$ for WASP-69\,b; $K_\mathrm{p} = 127$\,km\,s$^{-1}$ and $v_\mathrm{offset}=-6.8$\,km\,s$^{-1}$ for KELT-11\,b) and scaled with the fraction of the planetary disk intersecting with the stellar disk to mimic the changing signal strength during ingress and egress. The models were injected into the observed spectra before normalising and applying \texttt{SYSREM} as described in Sect. \ref{Sect_RemovingTellurics}. We estimated the detectability of the atmosphere in our observation by calculating the S/N in the $K_\mathrm{p}$-$v_\mathrm{offset}$ map at the injected position, following the procedure explained in the previous sections. 

To remove the random noise structure (and in the case of H$_2$O the tentative signal) at the injected position in the $K_\mathrm{p}$-$v_\mathrm{offset}$ map, we additionally calculated a CCF without an injected signal, which we subtracted from the injected CCF prior to calculating the $K_\mathrm{p}$-$v_\mathrm{offset}$ map. Then we used the standard deviation of the non-injected $K_\mathrm{p}$-$v_\mathrm{offset}$ map to convert the signal strength into units of S/N.

By repeating this injection-recovery test for a grid of abundances and cloud deck pressures, the boundary between detectable and non-detectable atmospheric features can be determined. We assumed a detection threshold of S/N$=5$. Hence, the combinations of abundances and cloud
deck pressures with a S/N > 5 in this injection
test would have been detected  in our observations, and
can therefore be ruled out.

\subsubsection{KELT-11\,b}
\begin{figure*}
\centering
\includegraphics[width=\hsize]{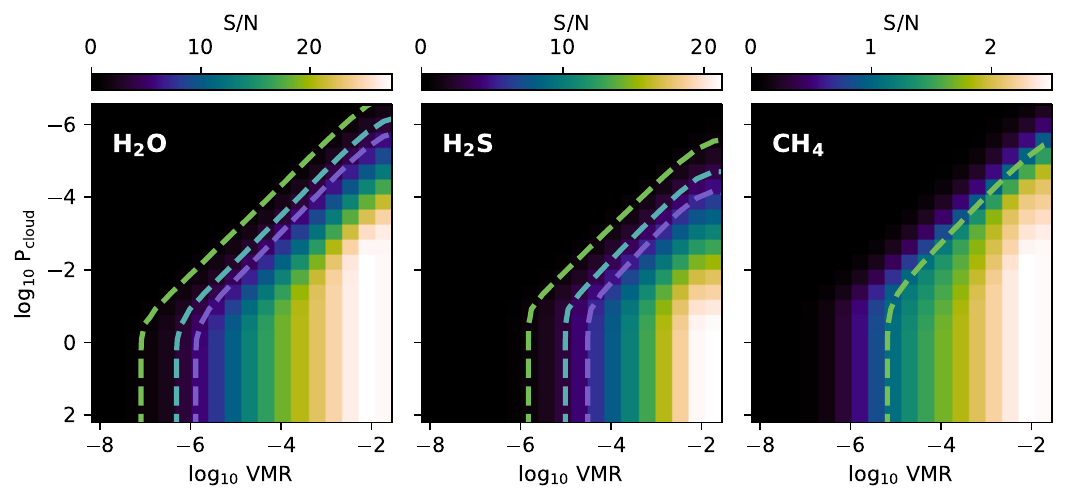}
  \caption{Results of the S/N grid analysis for  KELT-11\,b. Shown are the retrieved S/N values of injected synthetic models with varying molecular abundances and cloud deck pressures (in bar) for H$_2$O, H$_2$S, and CH$_4$ (note the different colour scales). The dashed lines indicate the thresholds corresponding to a S/N of 1 (green), 3 (teal), and 5 (purple).
  Notably, the S/N for CH$_4$ is consistently below 3, so only the S/N=1 threshold is shown.
          }
     \label{Fig_SNR_Grid_K11}
\end{figure*}

Figure \ref{Fig_SNR_Grid_K11} shows the S/N grid results for KELT-11\,b. A planetary H$_2$O signal of similar strength as the observed tentative signal could be explained by a parameter combination included in the region between the S/N = 3 and S/N = 5 boundaries. In the cloud-free case, this would correspond to a VMR of $10^{-6}$, while scenarios with clouds higher up in the atmosphere would require larger abundances.
Based on our non-detection of H$_2$S, we can rule out scenarios with VMRs of H$_2$S above $\sim 10^{-5}$ with simultaneous cloud deck pressures above $\sim 10^{-2}$\,bar. The strength of a potential CH$_4$ signal remains below a S/N of 3 even for very high abundances and cloud-free conditions. Therefore, we cannot place any constraints on the CH$_4$ abundance from our data.

\subsubsection{WASP-69\,b}
\begin{figure*}
\centering
\includegraphics[width=\hsize]{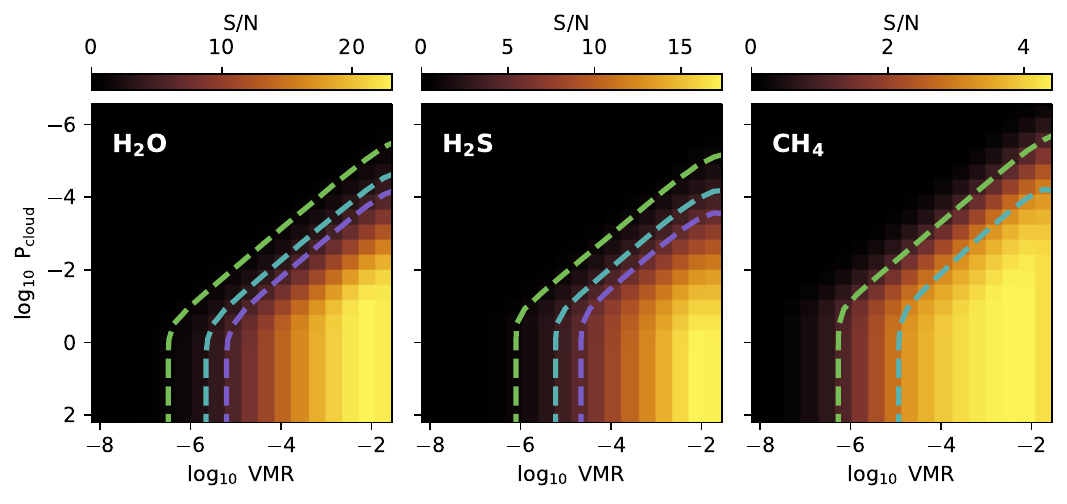}
  \caption{Same as Fig. \ref{Fig_SNR_Grid_K11} but for WASP-69\,b. 
          }
     \label{Fig_SNR_Grid_W69}
\end{figure*}

Figure \ref{Fig_SNR_Grid_W69} shows the S/N grids of H$_2$O, H$_2$S, and CH$_4$ for \mbox{WASP-69\,b}. We indicate the boundaries of S/N = 1, S/N = 3, and S/N = 5, and any parameter combinations with lower abundances or lower cloud deck pressures are consistent with our non-detections. Notably, clouds below a pressure of $\sim 1$\,bar have no major influence on the spectra, and in this scenario the abundances are constrained to $\log_{10}$H$_2$O$<-5.2$, $\log_{10}$H$_2$S$<-4.7$, according to the S/N = 5 boundary. In contrast, in the case of high clouds even large molecular abundances would not be detectable as the signal is muted to a high degree by the clouds.

The maximum S/N differs significantly between the three species. This is mainly due to the different number of spectral lines in the observed wavelength range, and their strength relative to the continuum (see the top panel in Fig. \ref{Fig_Models+Detmaps_W69}). Consequently, the constraint on the H$_2$O abundance is tighter than that on H$_2$S and CH$_4$. The signal strength of CH$_4$ does not cross the detection threshold even in the case of high abundances and effectively no cloud coverage. Therefore, we are not able to put any constraints on the CH$_4$ abundance from our observational data.

Under the hypothesis that the weak tentative H$_2$O signal in the $K_\mathrm{p}$-$v_\mathrm{offset}$ map is a planetary signal, we would expect a combination of H$_2$O abundance and cloud deck pressure corresponding roughly to the S/N = 3 boundary indicated in Fig. \ref{Fig_SNR_Grid_W69}. Alternatively, if this signal is due to random noise, the S/N=3 contour corresponds to the upper limit that is consistent with a non-detection.

\subsection{Atmospheric retrievals}

\begin{figure*}
\centering
\includegraphics[width=\hsize]{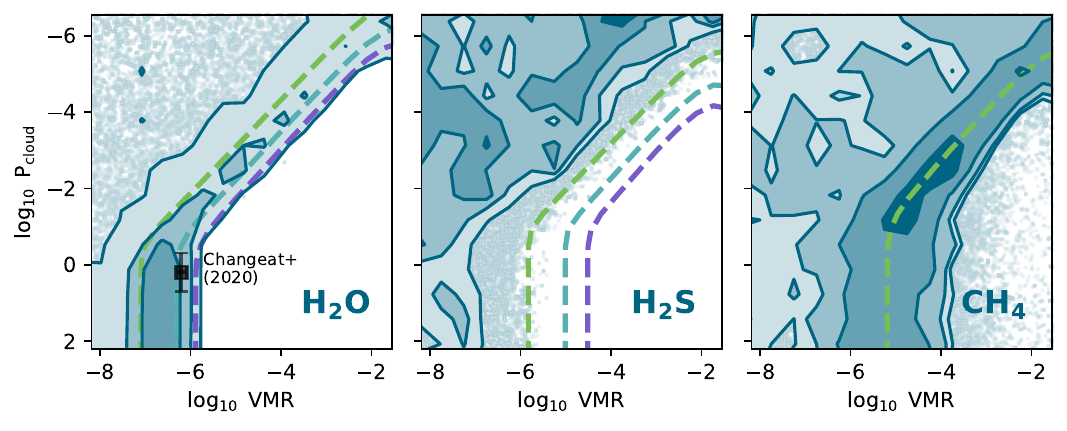}
  \caption{Joint posterior distributions of the molecular abundances and cloud deck pressure from atmospheric retrievals of KELT-11\,b. Each panel corresponds to a separate retrieval that includes only the indicated molecule as the opacity source. The dashed lines correspond to the S/N thresholds from the S/N grid analysis (Fig. \ref{Fig_SNR_Grid_K11}). 
          }
     \label{Fig_Retrieval_K11}
\end{figure*}

\begin{figure*}
\centering
\includegraphics[width=\hsize]{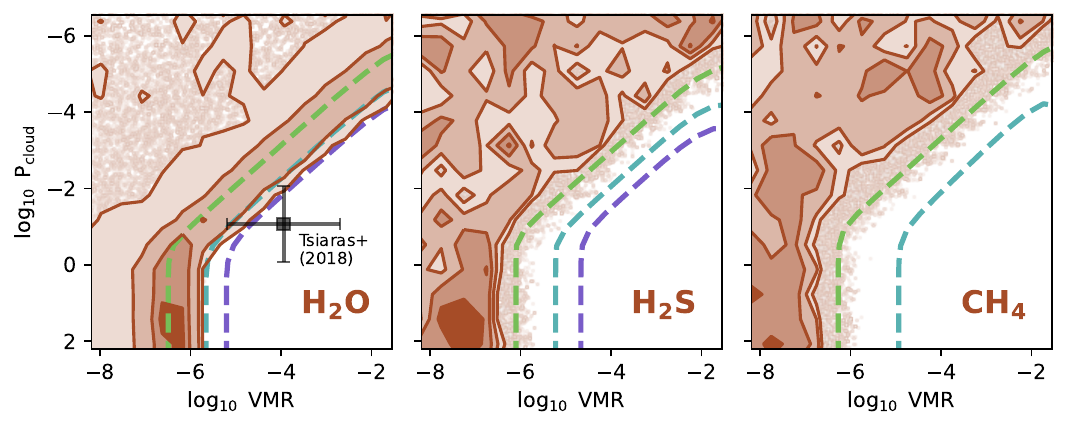}
  \caption{Same as Fig. \ref{Fig_Retrieval_K11} but for WASP-69\,b.
          }
     \label{Fig_Retrieval_W69}
\end{figure*}

Instead of comparing observed signal strengths with the results of injection-recovery tests, a common approach to determining molecular abundances involves atmospheric retrievals \citep[e.g.][]{brogi_retrieving_2019, gibson_detection_2020, maguire_high-resolution_2022, yan_crires_2023}. In the following, we use a Markov chain Monte Carlo (MCMC) retrieval framework to investigate which parameter combinations would be consistent with the observed tentative H$_2$O signals and non-detection of other species. Calculating a likelihood gives a measure of how well a given model fits to the data. Because we do not have a clear detection, the MCMC will not converge to a single parameter combination, and the posterior distribution will instead cover the entire region in the multi-dimensional parameter space that is consistent with our observations. For models with high abundances and high cloud deck pressure, the spectral lines are too deep to fit to the residual spectra, and the MCMC walkers will gravitate towards models with weaker line strengths.

\subsubsection{Retrieval framework} \label{Sect_RetrievalFramework}
We additionally performed individual retrievals for each species, in order to compare the results with the S/N grid method. A similar approach of applying a retrieval to non-detections was recently used by \citet{palle_exploring_2025} to constrain molecular abundances in the atmosphere of the rocky planet GJ 1132\,b. We calculated synthetic models with \texttt{petitRADTRANS}, which we then transformed into the same 3D format as the residual spectra (order $\times$ time $\times$ wavelength) and shifted into the planetary rest frame. We accounted for distortions introduced by the application of \texttt{SYSREM} following \citet{gibson_relative_2022}, calculating a filter matrix from the \texttt{SYSREM} vectors, which was applied to each individual model during the retrieval process.

We calculated the logarithmic likelihood function following \citet{hogg_data_2010} and \citet{yan_temperature_2020}:

\begin{align}
\ln(L) = - \frac{1}{2}\sum_{i,j}\left(\frac{(R_{i,j} - M_{i,j})^2}{(\beta \sigma_{i,j})^2} + \ln(2\pi(\beta \sigma_{i,j})^2)\right)\,,
\end{align}
where $R_{i,j}$ are the residual spectra at pixel $i$ and time $j$ with uncertainties $\sigma_{i,j}$, $M_{i,j}$ represents the 2D matrix of a filtered model spectrum, and $\beta$ is a scaling factor for the uncertainties. We sampled the parameter space with the MCMC algorithm as implemented in $\texttt{emcee}$ \citep{foreman-mackey_emcee_2013} with 32 walkers and 5\,000 steps each, discarding the first 1000 steps as burn-in.

The free parameters of the retrievals were the VMRs, the cloud deck pressure, and the scaling factor ($\beta$). We fixed $K_\mathrm{p}$ and $v_\mathrm{offset}$ to the values of the tentative H$_2$O signals. In all retrievals, the $T$-$p$ profile was fixed to an isothermal profile of the planet's respective equilibrium temperature, and no additional broadening (e.g. to account for winds) beyond the instrumental broadening was applied to the models.

\subsubsection{Retrieval results}
The posterior distributions of the retrievals are shown in Figs. \ref{Fig_Retrieval_K11} and \ref{Fig_Retrieval_W69} for KELT-11\,b and WASP-69\,b, respectively. 
The retrievals of all three investigated species do not converge to a clearly defined parameter region for either of the two planets, and the walkers instead cover a large part of the parameter space. This is expected in the absence of a clear planetary signal, and the covered regions represent the parameter combinations that are consistent with our non-detections. The boundaries of these regions generally agree well with the result from the S/N grid. The posterior distributions of the CH$_4$ retrieval for KELT-11\,b cover the entire parameter space and therefore we cannot place any constraints, which is consistent with the S/N grid result.

The scaling factor for the uncertainties, $\beta$, was found to be 0.70 for all retrievals of WASP-69\,b and 0.78 for all retrievals of KELT-11\,b. This indicates that the uncertainties determined by the reduction pipeline were slightly overestimated. Such a value of $\beta$ is in line with other retrieval studies of CARMENES data \citep[e.g.][]{blain_formally_2024}

\section{Discussion} \label{Sect_Discussion}

\subsection{KELT-11\,b}

The retrievals of KELT-11\,b do not converge and the abundance and cloud deck pressure are constrained to a broad region of the parameter space. The boundary of this region matches the S/N = 5 boundary of the S/N grid. Despite the tentative H$_2$O signal in the $K_\mathrm{p}$-$v_\mathrm{offset}$ map, the H$_2$O retrieval does not converge and results in similar upper limits as in the case of H$_2$S. This could potentially be an indication that the H$_2$O signal from the cross-correlation analysis is not a real planetary signal. If the signal was real, the blueshift of 6\,km\,s$^{-1}$ could hint towards the existence of a global day-to-night wind. In such a wind pattern, the atmospheric material at the terminator is moving towards the observer during the transit, inducing an overall blueshift in the transmission spectrum \citep[see e.g.][]{alonso-floriano_multiple_2019,sanchez-lopez_water_2019,prinoth_titanium_2022}. Further observations are required to definitively solve this question.

\citet{colon_unusual_2020} analysed the transmission spectrum of KELT-11\,b as observed with HST/WFC3 in combination with transit observations using \textit{Spitzer} \citep{beatty_determining_2017} and TESS. Their retrieval using all of these data and a full model with clouds and hazes results in a water abundance of $\log_{10}(\mathrm{H}_2\mathrm{O}) = -4.0^{+0.4}_{-0.5}$, while it is not able to place strong constraints on the presence of clouds. Such a water abundance would be consistent with our non-detection only in the presence of clouds above $\log_{10}(P_\mathrm{cloud}) \approx -3\,$bar. They note however that the water absorption feature observed with HST has an unusual shape, and the best-fitting models either require additional absorbers far out of chemical equilibrium or result in a poor fit to the data.
 
 \citet{changeat_kelt-11_2020} investigated the same datasets, and applied multiple different retrievals with varying absorbing species. Their `water-only retrieval', which is closest to the approach of this work, results in an atmosphere  with a low cloud height ($\log_{10}(P_\mathrm{cloud}\,[\mathrm{bar}]) = 0.2 \pm 0.5$) and a water abundance of $\log_{10}(\mathrm{H}_2\mathrm{O}) = -6.2 \pm 0.1$. This is in good agreement with the results of both our S/N grid and retrieval. The `base retrieval' scenario of \citet{changeat_kelt-11_2020} additionally contains carbon-bearing species and constrains the CH$_4$ abundance to a low value of $\log_{10}(\mathrm{CH}_4) = -9.7 ^{+1.7}_{-1.6}$. Our analysis shows that our observations cannot place any constraints on the CH$_4$ abundance, and abundances as low as the one retrieved from the HST observations are well beyond the detection capabilities of CARMENES, even when combining a large number of nights.

\subsection{WASP-69\,b}
Our cross-correlation analysis of the six observations of \mbox{WASP-69\,b} resulted in a S/N peak of H$_2$O close to the expected planetary rest frame. However, the strength of this peak is weak and there are other peaks of similar strength visible in the noise pattern of the $K_\mathrm{p}$-$v_\mathrm{offset}$ map. Additionally, the peak extends across a large range of $K_p$ values and includes the literature value of $K_{p,\mathrm{lit}}=127$\,km\,s$^{-1}$, but the maximum is reached at a smaller value of $K_p=89$\,km\,s$^{-1}$. Consequently, we cannot ensure the planetary origin of such a signal. \citet{guilluy_gaps_2022} observed three transits with GIANO-B (wavelength coverage 0.95 -- 2.45\,$\mu$m) and reported a detection of H$_2$O with a S/N of 4.1. Their cross-correlation framework results in a signal at $K_p = 114^{+59}_{-57}$\,km\,s$^{-1}$, while an alternative likelihood mapping approach peaks at a $K_p$ of $84^{+30}_{-32}$\,km\,s$^{-1}$, close to our value. Their signal has a similar extent in $K_p$ to that in our result.

\citet{tsiaras_population_2018} analysed a transit observed with HST and retrieved a VMR of $\log_{10}(\mathrm{H}_2\mathrm{O}) = -3.94 \pm 1.25$ and a cloud pressure of $\log_{10}(P_\mathrm{cloud}\,[\mathrm{bar}]) = -1.07 \pm 0.99$. According to our S/N grid analysis, such an atmosphere should have been detectable in our observations with a S/N of $\sim$10. Our upper limit on the H$_2$O abundance in this pressure range of $\log_{10}(\mathrm{H}_2\mathrm{O}) < -4.7$ falls below their value but is consistent within the uncertainties.
In contrast, the retrieval of FORS2 observations by \citet{petit_dit_de_la_roche_detection_2024} favoured models with high-altitude clouds. Compared to our results, their H$_2$O abundance and cloud deck pressure are higher, but the 1$\sigma$ region coincides with our S/N$=5$ contour in the S/N grid and the boundary of our retrieved posterior distribution.

Recent eclipse observations of WASP-69\,b with the JWST Near-InfraRed Camera (NIRCam) and Mid-Infrared Instrument (MIRI) showed absorption features of H$_2$O, CO, and CO$_2$ and no strong evidence of CH$_4$ \citep{schlawin_multiple_2024}. Their most plausible retrievals had to include high-altitude clouds ($P_\mathrm{cloud} \approx 10^{-5}\,$bar) and metallicities of up to 10 times the solar value in order to adequately fit the observations. These eclipse observations are sensitive to the planetary dayside atmosphere while our transit observations probe the terminator regions. Assuming that the cloud coverage does not differ significantly between these regions, any atmosphere with such a high-altitude cloud layer would not be detectable with our CARMENES observations regardless of the chemical composition.

Of our six observed nights, four nights suffered from severe telluric contamination because the radial velocity difference between planetary and telluric rest frame was very small during these transits (Fig. \ref{Fig_ObservationalConditions}). This hampers the ability to detect molecular species such as H$_2$O, which are present in both the planetary and telluric atmospheres. A further factor contributing to the non-detection was the increased number of \texttt{SYSREM} iterations required due to the severe telluric contamination, which may inadvertently remove part of the planetary signal.
In our analysis, we only excluded night 3 due to the persistent telluric contamination. Based on injection-recovery tests of H$_2$O, we decided to include all of the five remaining nights in our analysis.

\subsection{General remarks}
For the investigated species, the boundaries constrained by the S/N grid and the retrieval generally line up well with each other. In most cases,  the boundary established by the retrieval falls between the S/N = 3 and S/N = 1 boundaries of the S/N grid. Only for H$_2$O in WASP-69\,b, the retrieval is consistent with the S/N = 5 boundary. This difference could potentially be caused by the influence of a tentative planetary signal. Conversely, the S/N grid and the retrieval of KELT-11\,b agree well and the tentative H$_2$O signal has no apparent effect on these results.

In general, small deviations between the results of the two approaches can be expected due to the inherent differences. For the S/N grid calculation, we removed the noise structure in the $K_\mathrm{p}$-$v_\mathrm{offset}$ map before determining the S/N in order to mitigate the effect of the noise. However, the same cannot be done for the retrieval approach. In the case of CH$_4$ in WASP-69\,b, a large-scale negative feature can be seen in the $K_\mathrm{p}$-$v_\mathrm{offset}$ map at the expected planetary velocity. The retrieval can be sensitive to this anti-correlation between model and data, and subsequently result in a lower abundance threshold. Further, in the retrieval we applied the model filter of \citet{gibson_relative_2022} to account for distortions of the planetary signal due to \texttt{SYSREM}. In a future study, this filter could in principle be applied to the models used in the S/N grid, potentially reducing the discrepancies between the two approaches.

In our modelling framework, we applied several simplifying assumptions to reduce the model complexity. The molecular abundances and temperatures were assumed to be isobaric, and we fixed the temperatures to the equilibrium temperature values from the literature. In reality, the conditions of morning and evening terminator could vary in abundances, temperature and cloud properties \citep[e.g.][]{gandhi_spatially_2022, nortmann_crires_2025}. In addition, a significant super-rotation due to an equatorial jet could separate the contributions from the two terminators in velocity space, which could reduce the signal strength or hide the signal completely when using models without this effect \citep{nortmann_crires_2025}.

In this work, we treated each molecular species individually, while in reality the atmosphere consists of a mixture of many species. Strong absorption by a species with a dense forest of lines (e.g. H$_2$O) can hide contributions from other species, especially in the presence of a cloud deck. For the S/N grid, the simultaneous inclusion of multiple species, each with its abundance as a free parameter, is computationally not feasible. However, the same limitation does not apply to the Bayesian retrieval approach, which can feasibly explore a higher-dimensional parameter space. In order to test whether the inclusion of multiple species influences the retrieved constraints, we conducted additional retrievals of both planets that simultaneously includes H$_2$O, H$_2$S, and CH$_4$ (Appendix \ref{Sect_CombinedRetrieval}). The resulting posterior distributions of the molecular abundances and cloud deck pressure are very similar to the previous results using individual retrievals with only a single species. For KELT-11\,b, the constraints on H$_2$O were slightly weaker with a median difference of 0.67\,dex. The combined retrieval reinforced the conclusion that CH$_4$ is unconstrained by our data. For WASP-69\,b, the abundances of H$_2$S, and CH$_4$ were slightly weaker constrained in scenarios with very low cloud deck pressures. In the region of high altitude clouds ($\log_{10}(P_\mathrm{cloud}\,[\mathrm{bar}]) < -3$), the median difference between the retrieval versions was 0.01\,dex for H$_2$O, 0.16\,dex for H$_2$S, and 0.45\,dex for CH$_4$. We attributed these differences to the influence of absorption lines of different species overlapping with each other, as well as to slight changes in the overall mean molecular weight resulting in less prominent spectral features. Especially in scenarios in which the atmospheric signal is already muted by a high cloud deck, the overlap of spectral lines can result in weaker constraints as even higher molecular abundances are in agreement with the non-detection in our data.  Because we did not find a significant difference between the individual and the combined retrieval, we opted to apply the same approach of individually analysing each species for both the retrieval and the S/N grid to facilitate the comparison between the two methods.

\section{Summary} \label{Section_Conclusion}
We analysed the transmission spectra of KELT-11\,b and \mbox{WASP-69\,b} using multiple high-resolution spectroscopic observations with CARMENES. Using the cross-correlation method, we found a tentative signal of H$_2$O in KELT-11\,b, while searches for additional molecular species yielded non-detections. We applied two complementary approaches to constrain the atmospheric abundances and cloud deck pressures that are consistent with these findings. We used a S/N grid based on injection-recovery tests to determine the theoretical cross-correlation signal strength of models with different parameter combinations. As an alternative method, a Bayesian retrieval was used to identify regions in the parameters space consistent with our observational data. The two methods yielded similar constraints, although there are nuanced differences for some of the species. Our results for WASP-69\,b agree with previous ground-based studies, but they are not consistent with abundances retrieved from HST observations. Conversely, the upper limits on the H$_2$O abundance of KELT-11\,b align with the existing literature. The retrieval approach could be used in the future to simultaneously determine constraints on multiple species.

\begin{acknowledgements}
CARMENES is an instrument at the Centro Astron\'omico Hispano en Andaluc\'ia (CAHA) at Calar Alto (Almer\'{\i}a, Spain), operated jointly by the Junta de Andaluc\'ia and the Instituto de Astrof\'isica de Andaluc\'ia (CSIC).
 
CARMENES was funded by the Max-Planck-Gesellschaft (MPG), 
the Consejo Superior de Investigaciones Cient\'{\i}ficas (CSIC),
the Ministerio de Econom\'ia y Competitividad (MINECO) and the European Regional Development Fund (ERDF) through projects FICTS-2011-02, ICTS-2017-07-CAHA-4, and CAHA16-CE-3978, 
and the members of the CARMENES Consortium 
(Max-Planck-Institut f\"ur Astronomie,
Instituto de Astrof\'{\i}sica de Andaluc\'{\i}a,
Landessternwarte K\"onigstuhl,
Institut de Ci\`encies de l'Espai,
Institut f\"ur Astrophysik G\"ottingen,
Universidad Complutense de Madrid,
Th\"uringer Landessternwarte Tautenburg,
Instituto de Astrof\'{\i}sica de Canarias,
Hamburger Sternwarte,
Centro de Astrobiolog\'{\i}a and
Centro Astron\'omico Hispano-Alem\'an), 
with additional contributions by the MINECO, 
the Deutsche Forschungsgemeinschaft (DFG) through the Major Research Instrumentation Programme and Research Unit FOR2544 ``Blue Planets around Red Stars'', 
the Klaus Tschira Stiftung, 
the states of Baden-W\"urttemberg and Niedersachsen, 
and by the Junta de Andaluc\'{\i}a.

We acknowledge financial support from the DFG through Germany’s Excellence Strategy EXC 2094 – 39078331 and project 314665159; the LMU-Munich Fraunhofer-Schwarzschild Fellowship; the Agencia Estatal de Investigaci\'on (AEI/10.13039/501100011033) of the Ministerio de Ciencia e Innovaci\'on and the ERDF ``A way of making Europe'' through projects 
  PID2022-137241NB-C4[1:4],     
  PID2021-125627OB-C31,         
  PID2022-141216NB-I00,
and the Centre of Excellence ``Severo Ochoa'' and ``Mar\'ia de Maeztu'' awards to the Instituto de Astrof\'isica de Canarias (CEX2019-000920-S), Instituto de Astrof\'isica de Andaluc\'ia (CEX2021-001131-S) and Institut de Ci\`encies de l'Espai (CEX2020-001058-M). This work was co-funded by the European Union (ERC-CoG, EVAPORATOR, Grant agreement No. 101170037). Views and opinions expressed are however those of the author(s) only and do not necessarily reflect those of the European Union or the European Research Council. Neither the European Union nor the granting authority can be held responsible for them.

\end{acknowledgements}

%
%
\bibliographystyle{aa} 
\bibliography{references_new}

\newpage
\begin{appendix}
\onecolumn

\section{Observational conditions}
\begin{center}
   \includegraphics[width=0.8\textwidth]{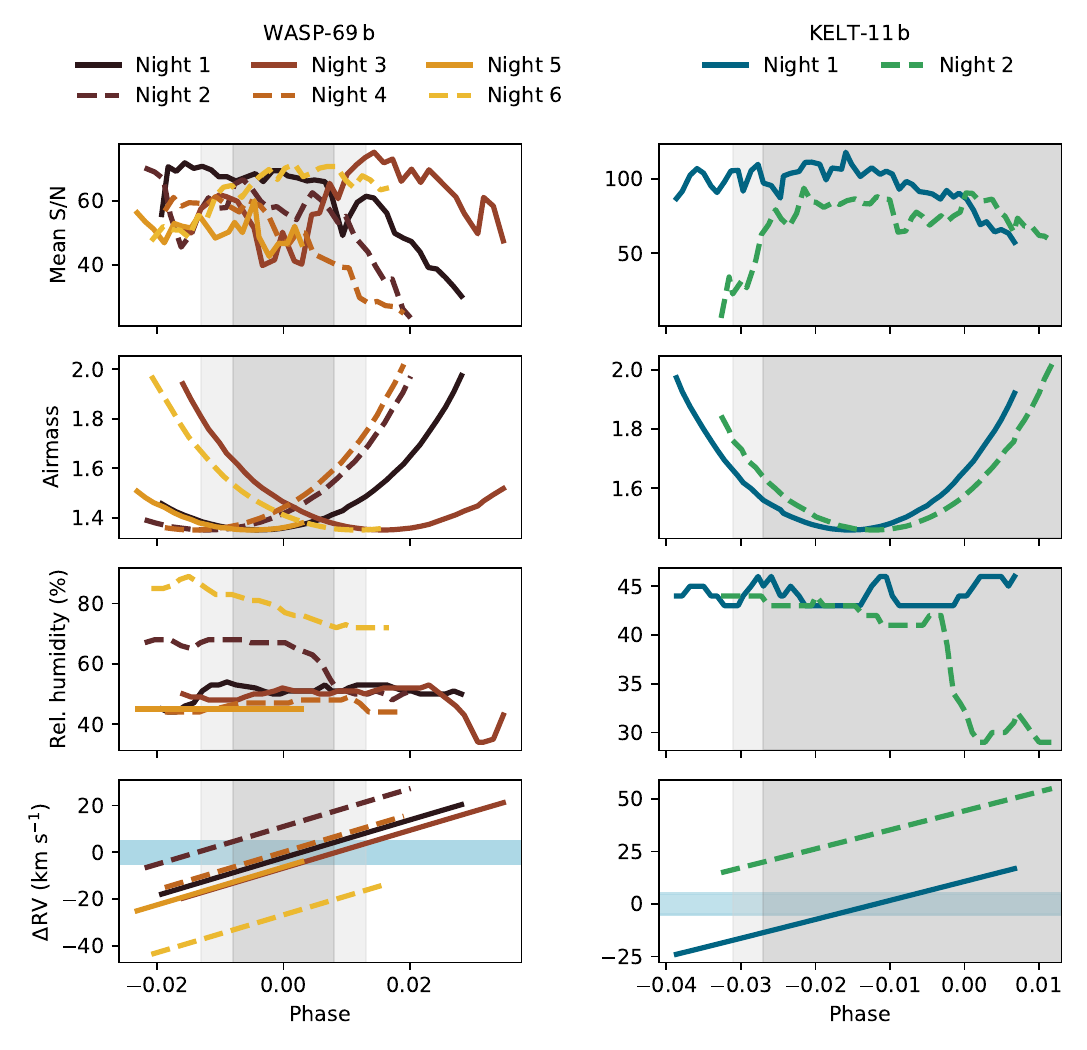}
  \captionof{figure}{Observational conditions and radial velocity shift between the telluric and planetary rest frame for the six nights of WASP-69\,b and two nights of KELT-11\,b observations. The vertical light shaded regions indicate the ingress and egress, and the dark shaded region shows the transit. The horizontal blue regions in the lower panels show the regions corresponding to a shift between $-5$\,km\,s$^{-1}$ and $+5$\,km\,s$^{-1}$. In this region, the observation is particularly affected by telluric contamination.}
  \label{Fig_ObservationalConditions}
\end{center}

\section{Cross-correlation functions}

\begin{center}
   \includegraphics[width=\textwidth]{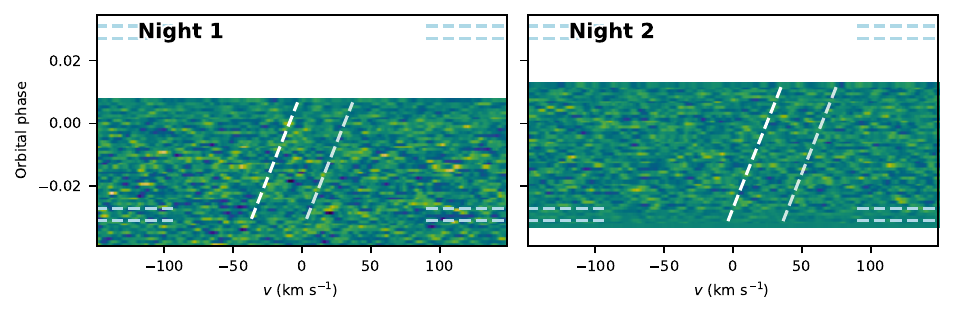}
  \captionof{figure}{CCFs of H$_2$O for the two nights of KELT-11\,b. For each night, the shown CCF corresponds to the optimised number of \texttt{SYSREM} iterations as detailed in Table \ref{table:Sysrem iterations}. The planetary rest frame follows a trail between the dashed white lines. The horizontal dashed lines indicate the contact points (T1, T2, T3, and T4, from bottom to top).
   Due to the long transit duration of KELT-11\,b, the end of the transit was not covered on either of the two nights.}
  \label{Fig_CCF_K11}
\end{center}

\begin{center}
   \includegraphics[width=\textwidth]{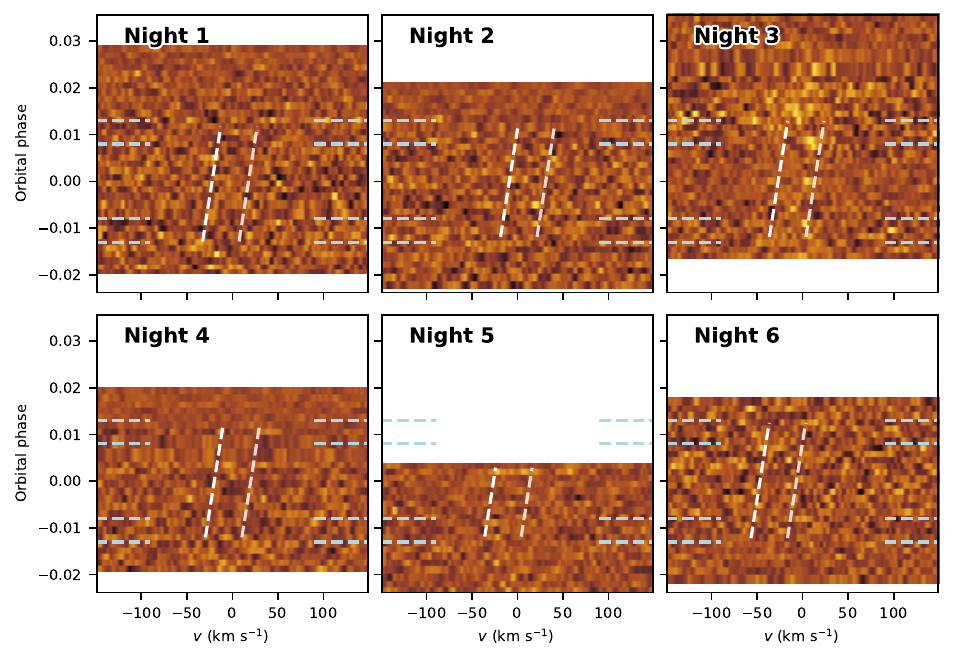}
  \captionof{figure}{Same as Fig. \ref{Fig_CCF_K11} but for WASP-69\,b. Each night covers a slightly different range of phases, and in particular night 5 only includes the first half of the transit. Shown are the CCFs after seven iterations for all nights.}
  \label{Fig_CCF_W69}
\end{center}

\newpage
\section{Individual $K_\mathrm{p}$-$v_\mathrm{offset}$ maps for H$_2$O in KELT-11\,b}
\begin{center}
   \includegraphics[width=0.8\textwidth]{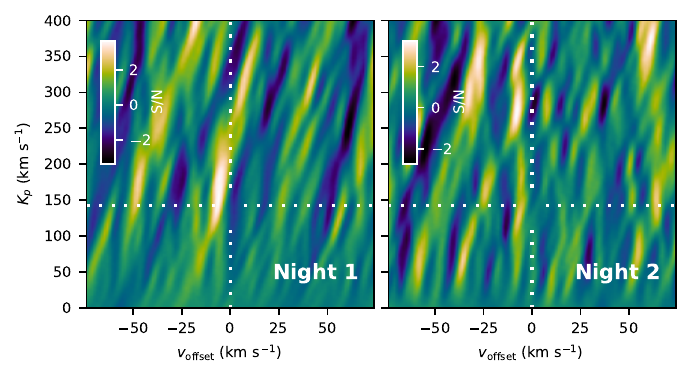}
  \captionof{figure}{$K_\mathrm{p}$-$v_\mathrm{offset}$ maps of H$_2$O for KELT-11\,b, showing the contributions from the two individual nights. The dotted lines indicate the expected location of the planetary signal.}
  \label{Fig_Detmap_K11_IndividualNights}
\end{center}

\section{H$_2$O in WASP-69\,b using optimised \texttt{SYSREM} iterations}

\begin{center}
   \includegraphics[width=\textwidth]{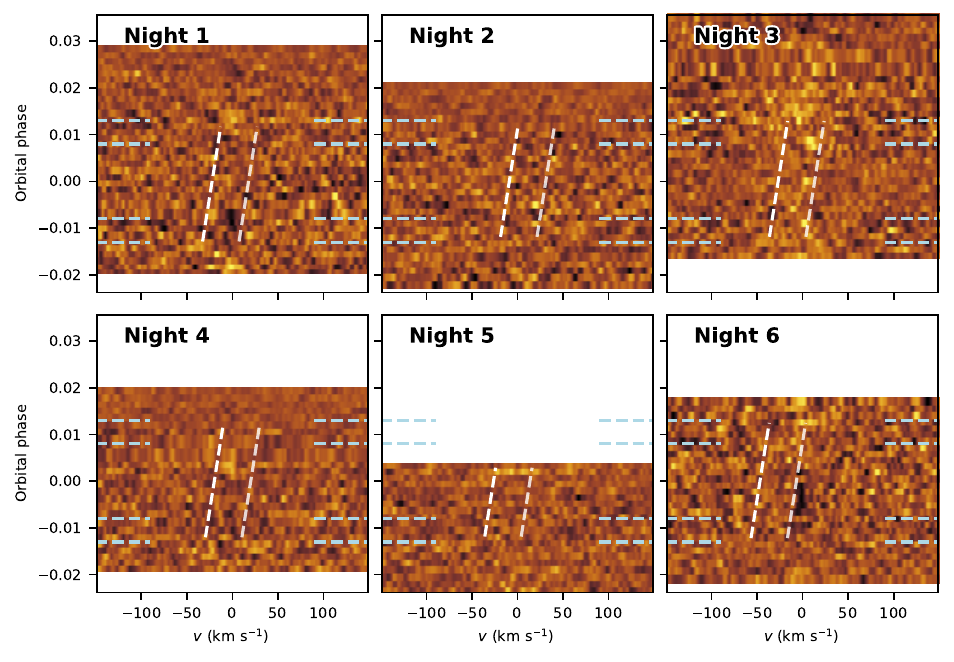}
  \captionof{figure}{CCFs of H$_2$O for the six nights of WASP-69\,b, using the optimised number of \texttt{SYSREM} iterations (3, 2, 6, 3, 2, and 2 iterations for nights 1 through 6, respectively). The planetary rest frame follows a trail between the dashed lines. The horizontal dashed lines indicate the contact points (T1, T2, T3, and T4, from bottom to top). Strong telluric residuals located at $v=0$\,km\,s$^{-1}$ are visible for some of the nights.}
  \label{Fig_CCF_W69_optimised}
\end{center}

\begin{center}
   \includegraphics[width=0.35\textwidth]{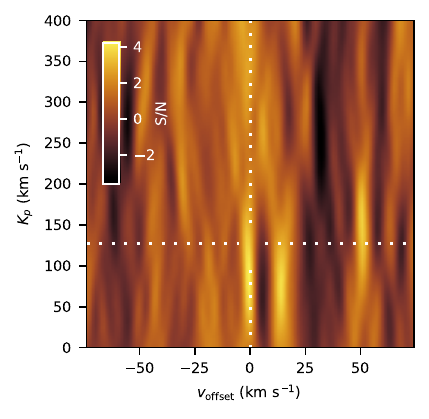}
  \captionof{figure}{$K_\mathrm{p}$-$v_\mathrm{offset}$ maps of H$_2$O for WASP-69\,b, calculated from the CCFs with an optimised number of \texttt{SYSREM} iterations shown in Fig. \ref{Fig_CCF_W69_optimised}. The dotted lines indicate the expected location of the planetary signal.}
  \label{Fig_Detmap_W69_H2O_OptimisedSYSREM}
\end{center}

\section{Combined retrieval with all species} \label{Sect_CombinedRetrieval}
We conducted retrievals for both planets that simultaneously included H$_2$O, H$_2$S, and CH$_4$, while otherwise following the set-up as described in Sect. \ref{Sect_RetrievalFramework}. We applied the same number of  \texttt{SYSREM} iterations as in the individual retrievals of H$_2$O. The abundances of the three species were set as free, independent parameters, and they were assumed to be constant with pressure. In order to generate a self-consistent model spectrum, the mean molecular weight was calculated based on these abundances during each model calculation.
The full corner plots of these combined retrievals are shown in Figs. \ref{Fig_Cornerplot_CombinedRetrieval_full_K11} and \ref{Fig_Cornerplot_CombinedRetrieval_full_W69}, while Figs. \ref{Fig_Retrieval_K11_CombinedRetrieval} and \ref{Fig_Retrieval_W69_CombinedRetrieval} compare the posterior distributions of abundances and cloud deck pressure from the combined retrievals with the results from the three individual retrievals shown in Figs. \ref{Fig_Retrieval_K11} and \ref{Fig_Retrieval_W69}, respectively.

\begin{center}
   \includegraphics[width=0.7\textwidth]{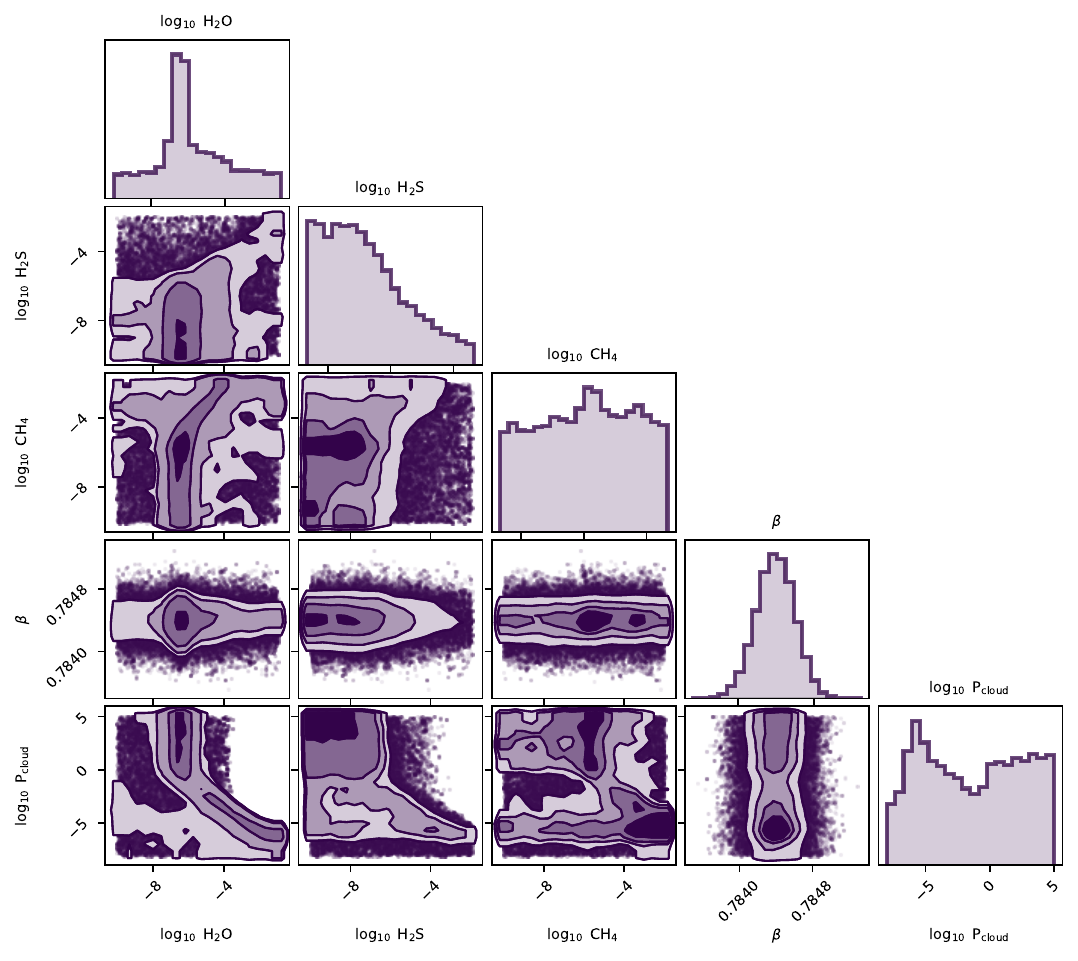}
  \captionof{figure}{Full corner plot of the posterior distributions from the combined retrieval of KELT-11\,b with H$_2$O, H$_2$S, and CH$_4$.}
  \label{Fig_Cornerplot_CombinedRetrieval_full_K11}
\end{center}

\begin{center}
   \includegraphics[width=0.7\textwidth]{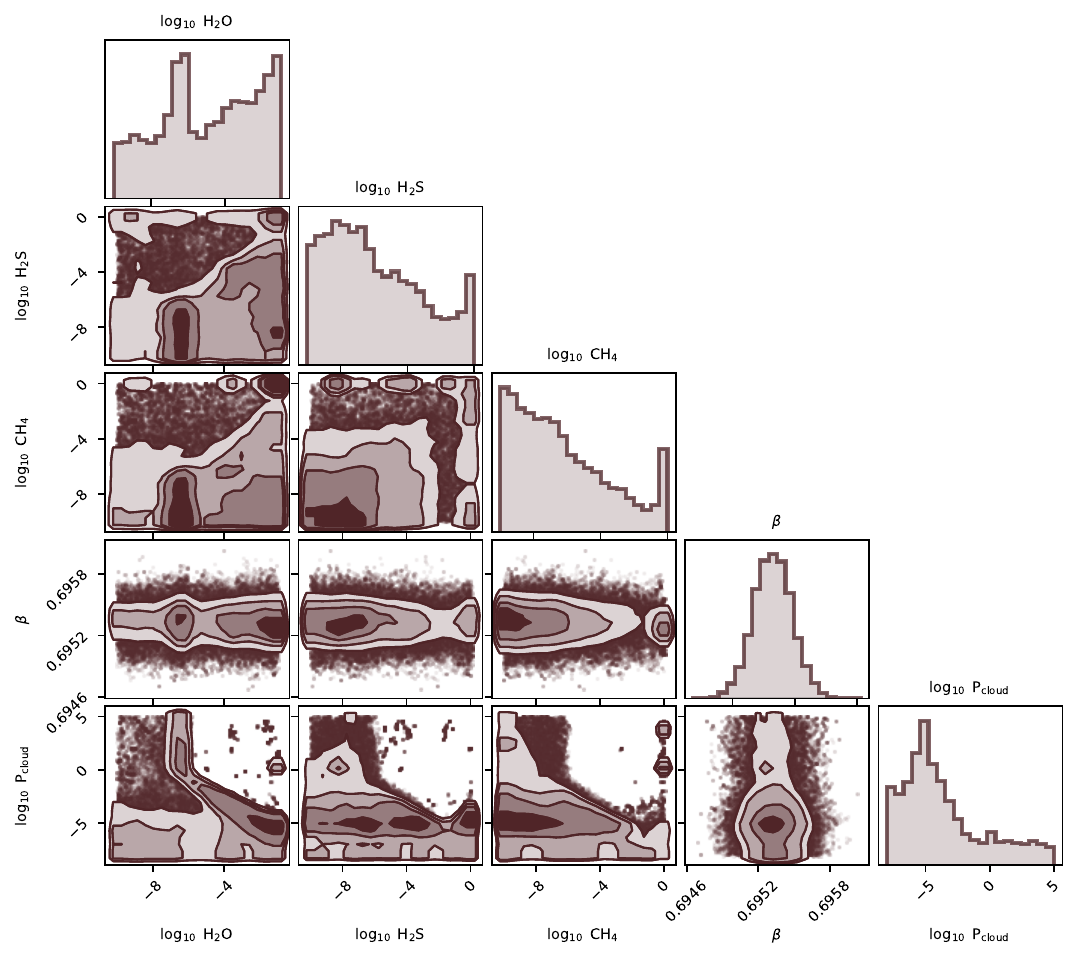}
  \captionof{figure}{Full corner plot of the posterior distributions from the combined retrieval of WASP-69\,b with H$_2$O, H$_2$S, and CH$_4$.}
  \label{Fig_Cornerplot_CombinedRetrieval_full_W69}
\end{center}

\begin{center}
   \includegraphics[width=0.8\textwidth]{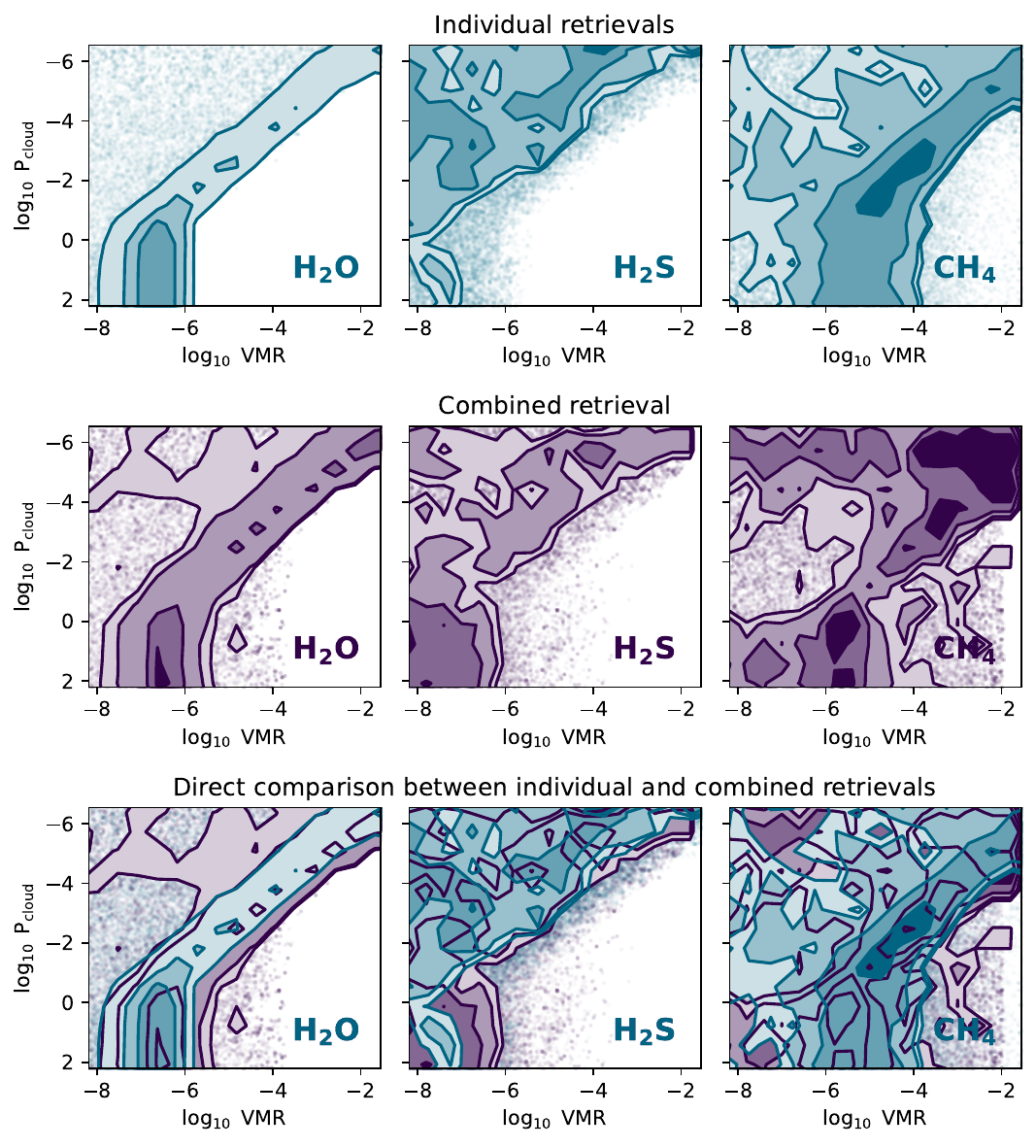}
  \captionof{figure}{Joint posterior distributions of the molecular abundances and cloud deck pressure from the individual retrievals of KELT-11\,b (top row), the combined retrieval including all three species (middle row), and an overlay of the two results for a direct comparison (bottom row).}
  \label{Fig_Retrieval_K11_CombinedRetrieval}
\end{center}

\begin{center}
   \includegraphics[width=0.8\textwidth]{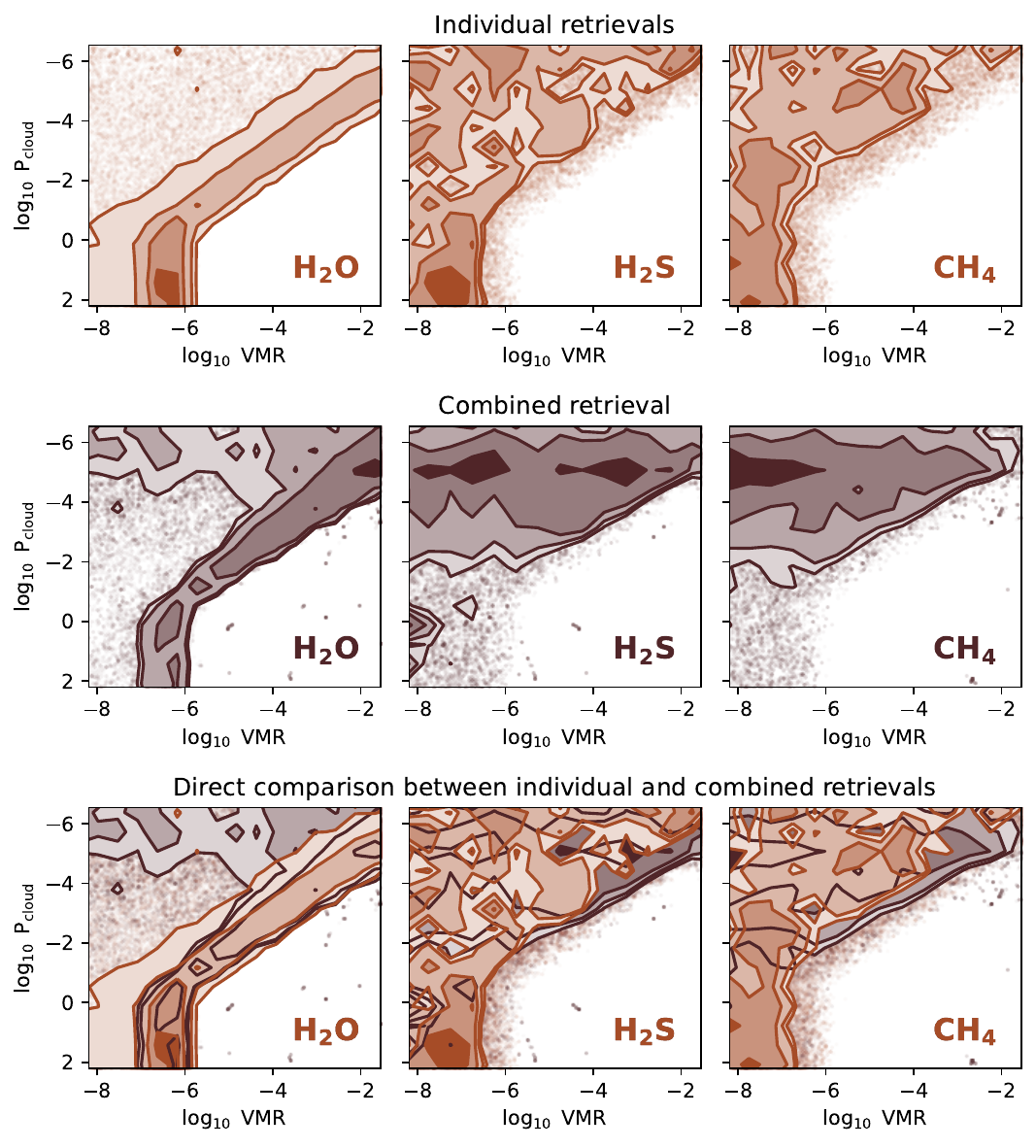}
  \captionof{figure}{Joint posterior distributions of the molecular abundances and cloud deck pressure from the individual retrievals of WASP-69\,b (top row), the combined retrieval including all three species (middle row), and an overlay of the two results for a direct comparison (bottom row).}
  \label{Fig_Retrieval_W69_CombinedRetrieval}
\end{center}

\end{appendix}
\end{document}